\documentclass[useams,usegraphicx]{mn2e}

\title{Two populations of metal-free stars in the early Universe}

\author[T. H. Greif and V. Bromm]{Thomas H. Greif\thanks{E-mail: tgreif@astro.as.utexas.edu} and Volker Bromm\\Department of Astronomy, University of Texas, Austin, TX 78712, USA}

\begin{document}

\maketitle

\topmargin-1cm

\begin{abstract}
We construct star formation histories at redshifts $z\ga 5$ for two physically distinct populations of primordial, metal-free stars, motivated by theoretical and observational arguments that have hinted towards the existence of an intermediate stellar generation between Population~III and Population~I/II. Taking into account the cosmological parameters as recently revised by the {\it Wilkinson Microwave Anisotropy Probe} after three years of operation, we determine self-consistent reionization histories and discuss the resulting chemical enrichment from these early stellar generations. We find that the bulk of ionizing photons and heavy elements produced at high redshifts must have originated in Population~II.5 stars, which formed out of primordial gas in haloes with virial temperatures $\ga 10^4~\rmn{K}$, and had typical masses $\ga 10~\rmn{M}_{\odot}$. Classical Population~III stars, formed in minihaloes and having masses $\ga 100~\rmn{M}_{\odot}$, on the other hand, had only a minor impact on reionization and early metal enrichment. Specifically, we conclude that only $\simeq 10$~per~cent by mass of metal-free star formation went into Population~III.
\end{abstract}

\begin{keywords}
stars: early-type -- stars: formation -- galaxies: formation -- galaxies: high-redshift -- cosmology: theory -- early Universe.
\end{keywords}

\section{Introduction}
Understanding the formation and properties of the first stars, formed out of pure H/He gas at the end of the cosmic dark ages, is one of the key problems in modern cosmology (e.g. Barkana \& Loeb 2001; Bromm \& Larson 2004). Within the standard $\Lambda$ cold dark matter ($\Lambda$CDM) model of structure formation (e.g. Spergel et al. 2006), the first stars are predicted to have formed in `minihaloes' of total mass $\ga 10^{6}~\rmn{M}_{\odot}$ at redshifts $z\ga 20$ (Haiman, Thoul \& Loeb 1996; Tegmark et al. 1997; Yoshida et al. 2003). Their emergence signalled a fundamental transition in cosmic history, from an initial state of simplicity to one of increasing complexity (e.g. Ciardi \& Ferrara 2005). Specifically, the first stars enriched the intergalactic medium (IGM) with the first heavy elements (Gnedin \& Ostriker 1997; Madau, Ferrara \& Rees 2001; Mackey, Bromm \& Hernquist 2003; Furlanetto \& Loeb 2003, 2005; Scannapieco, Schneider \& Ferrara 2003; Bromm, Yoshida \& Hernquist 2003; Yoshida, Bromm \& Hernquist 2004; Schneider et al. 2006). Secondly, they contributed to the reionization of the Universe, as inferred from the optical depth to Thomson scattering measured by the {\it Wilkinson Microwave Anisotropy Probe} ({\it WMAP}; Spergel et al. 2006), due to their efficient production of ionizing photons. To elucidate these feedback effects, we have to ask: What were the properties of the first generations of metal-free stars?

Numerical simulations imply that star formation inside `minihaloes' relied on molecular hydrogen cooling, and resulted in typically very massive so-called Population~III (Pop~III) stars with characteristic masses $\ga 100~\rmn{M}_{\odot}$ (Bromm, Coppi \& Larson 1999, 2002; Abel, Bryan \& Norman 2000, 2002; Nakamura \& Umemura 2001). Minihaloes are characterized by virial temperatures below $10^{4}$~K, and the star forming gas did not go through an ionized phase, prior to the late stages of the protostellar accretion process. In environments, however, where such ionization did occur, the thermal evolution of the primordial gas is predicted to have been markedly different (see Johnson \& Bromm 2006a,b for details). This difference arises because the hydrogen deuteride (HD) molecule provides an additional cooling channel, allowing the metal-free gas to cool to temperatures below those possible solely via H$_2$ cooling (e.g. Flower et al. 2000; Galli \& Palla 2002; Nakamura \& Umemura 2002). Gas fragmentation in cases where HD cooling is important could therefore have resulted in metal-free stars that were somewhat less massive than `classical' Pop~III stars. These stars, having typical masses $\ga 10~\rmn{M}_{\odot}$, have been termed Population~II.5 (Pop~II.5), indicating their transitional status in terms of typical mass and formation epoch between Pop~III and Pop~I/II (Mackey et al. 2003; Johnson \& Bromm 2006a). Connected to this, Tumlinson, Venkatesan \& Shull (2004) and Tumlinson (2006a,b), together with other groups, have argued that the observed chemical abundances in metal-poor Galactic halo stars are best explained by assuming that the first stars were somewhat less massive, different from the theoretical expectation (see Section~4). Our model could provide a natural explanation for such an intermediate mass scale, both in terms of the physics of primordial gas cooling and in terms of the cosmological environment where such stars could have formed.

The conditions necessary for HD cooling to become important were possibly met in a number of high-$z$ environments (Johnson \& Bromm 2006a). Among them are fossil H\,{\sc ii} regions, left behind after the death of a Pop~III star (e.g. Nagakura \& Omukai 2005; Susa \& Umemura 2006; Johnson \& Bromm 2006a,b; Abel, Wise \& Bryan 2006), and shocks induced either by the first supernova (SN) explosions (e.g. Machida et al. 2005), or as a result of hierarchical structure formation during the assembly of the first dwarf galaxies (e.g. Uehara \& Inutsuka 2000; Nakamura \& Umemura 2002). In this paper, we consider in particular the latter possibility, assuming that Pop~II.5 stars formed during the virialization of primordial haloes with virial temperatures $\ga 10^4$~K, which is sufficient to ionize the collapsing gas.

In the following, we construct separate cosmic star formation histories for the two distinct populations of metal-free stars (Section~2). We then compare the ionizing photon and heavy element production of our model to observations (Section~3), and conclude by summarizing our results and discussing the relevance of our work in light of recent developments (Section~4).

Throughout this paper, we adopt a $\Lambda$CDM model of hierarchical structure formation, with cosmological parameters consistent with the {\it WMAP} 3-yr results (Spergel et al. 2006), i.e. density in matter $\Omega_{\rmn{m}}=1-\Omega_{\Lambda}=0.24$, density in baryons $\Omega_{\rmn{b}}=0.042$, Hubble parameter $h=H_{0}/\left(100~\rmn{km}~\rmn{s}^{-1}~\rmn{Mpc}^{-1}\right)=0.73$, tilt $n_{\rmn{s}}=0.95$ and a normalization of the power spectrum to $\sigma_{8}=0.74$. For calculations involving the latter, we use the transfer function of Eisenstein \& Hu (1999).

\section{Cosmic Star Formation History}
\subsection{Population~III}
As the very first generation of stars to form out of primordial gas, Pop~III has been investigated extensively over the last decade (e.g. Barkana \& Loeb 2001; Bromm \& Larson 2004; Ciardi \& Ferrara 2005; Glover 2005). Already prior to the first numerical simulations of early structure formation, molecular hydrogen was identified as the primary coolant for pure H/He gas (e.g. Peebles \& Dicke 1968; Couchman \& Rees 1986; Tegmark et al. 1997). This view has been confirmed by detailed simulations since 1996 (e.g. Haiman et al. 1996; Abel et al. 1998, 2000; Bromm et al. 1999, 2002; Fuller \& Couchman 2000; Nakamura \& Umemura 2001; Yoshida et al. 2003, 2006). Within variants of the CDM model, the first stars are predicted to have formed in minihaloes of total mass $\simeq 10^6~\rmn{M}_{\odot}$ at redshifts $z\ga 20$. Such minihaloes have virial temperatures in excess of a minimum value, $T_{\rmn{min}}\simeq 10^{3}~\rmn{K}$, required for collapsing gas to cool in less than a Hubble time (e.g. Haiman et al. 1996; Tegmark et al. 1997, Yoshida et al. 2003). The evolution of gas inside minihaloes, driven by H$_2$ cooling, leads to the formation of protostellar fragments with typical masses $\ga 100~\rmn{M}_{\odot}$ (Bromm et al. 1999, 2002; Abel et al. 2000, 2002; Yoshida et al. 2006). We adopt this characteristic mass scale, and assume throughout this paper that Pop~III has an extremely top-heavy initial mass function (IMF) with cutoffs at $100$ and $500~\rmn{M}_{\odot}$, and a standard Salpeter (1955) slope. Consistent with simulation results (Abel et al. 2000, 2002; Bromm et al. 2002) and recent semianalytic treatments predicting a low Pop~III star formation efficiency (e.g. Haiman \& Bryan 2006; Choudhury \& Ferrara 2006), we assume $f_{*}\sim 0.001$, which corresponds to $\ga 1$ stars per halo. However, since uncertainties remain, we will reconsider our choice of this crucial quantity in later sections.

\subsubsection{Population~III Star Formation}
A straigthforward semianalytic approach to determine the star formation rate (SFR) for Pop~III is given by the Press-Schechter (PS) formalism (Press \& Schechter 1974), providing an expression for the abundance of dark matter haloes. An improved fit to simulation results is obtained by assuming ellipsoidal instead of spherical collapse, introduced by Sheth, Mo \& Tormen (2001). In quantitative studies (e.g. Heitmann et al. 2006; Reed et al. 2006), the Sheth-Tormen (ST) approach has proven to be significantly more accurate, with maximum deviations of $\simeq 50$~per~cent, leading us to adopt this formalism for all further calculations. By comparison, the PS mass function underpredicts the abundance of haloes at very high redshifts by almost an order of magnitude, although our qualitative results remain unaffected.

Using the ST approximation, we can estimate the number of dark matter haloes, $N(M,z)$, per unit mass and comoving volume at a given redshift:
\begin{equation}
\frac{\rmn{d}N(M,z)}{\rmn{d}M\rmn{d}V}=n_{\rmn{ST}}(M,z)\mbox{\ .}
\end{equation}
Since only a small fraction of haloes is massive enough to fulfil the cooling condition imposed by $T_{\rmn{vir}}>T_{\rmn{min}}=10^{3}~\rmn{K}$, we must adjust the limits of integration appropriately when calculating the collapsed fraction of mass, $F_{\rmn{col}}(z)$, available for Pop~III star formation:
\begin{equation}
F_{\rmn{col}}(z)=\frac{1}{\rho_{\rmn{m}}}\int_{M_{\rmn{min}}}^{M_{\rmn{max}}}\rmn{d}MMn_{\rmn{ST}}(M,z)\mbox{\ ,}
\end{equation}
where $\rho_{\rmn{m}}$ is the total mass density of the background universe. The limits in the integral, $M_{\rmn{min}}=M_{\rmn{crit}}(T_{\rmn{min}})$ and $M_{\rmn{max}}=M_{\rmn{crit}}(T_{\rmn{max}})$, are related to the virial temperature by:
\begin{equation}
M_{\rmn{crit}}=10^{8}~\rmn{M}_{\odot}\left(\frac{\mu}{0.6}\right)^{-3/2}\left(\frac{T_{\rmn{vir}}}{10^{4}~\rmn{K}}\right)^{3/2}\left(\frac{1+z}{10}\right)^{-3/2}\mbox{\ ,}
\end{equation}
where $\mu$ is the mean molecular weight, i.e. $1.2$ for neutral and $0.6$ for ionized, primordial gas. Since the Universe is mostly neutral at high redshifts, we choose $\mu =1.2$ for all further calculations. The lower mass limit in the integral corresponds to the minimum virial temperature below which haloes cannot cool efficiently. We set the upper limit to $T_{\rmn{max}}=10^{4}~\rmn{K}$, as haloes with masses $\ga 10^{8}~\rmn{M}_{\odot} $ go through a more complicated cooling process, leading to the formation of stars with characteristic masses $\ga 10~\rmn{M}_{\odot}$, termed Pop~II.5.  This novel generation of stars will be discussed in Section~2.2.

Applying the above criteria, we obtain a preliminary form for the Pop~III SFR:
\begin{equation}
\Psi_{*}(z)=\rho_{\rmn{m}}\frac{\Omega_{\rmn{b}}}{\Omega_{\rmn{m}}}f_{*}\left|\frac{\rmn{d}F_{\rmn{col}}(z)}{\rmn{d}z}\right|\left|\frac{\rmn{d}z}{\rmn{d}t}\right|\mbox{\ .}
\end{equation}
The solid line in Panel~(b) of Fig.~1 shows the result, implying that Pop~III star formation peaks around $z\simeq 10$, and declines sharply towards $z\simeq 5$, as an increasing number of minihaloes merge into larger haloes with masses above $M_{\rmn{max}}$. We emphasize that this is {\it not} the final SFR, as we have neglected all forms of feedback. In the next few sections, we will therefore investigate the impact of relevant feedback mechanisms on Pop~III.

\subsubsection{Feedback Mechanisms}
At some point during primordial star formation, negative feedback effects must have become important, eventually terminating the Pop~III mode, as stars in the present-day Universe are characterized by much higher metallicities and typically have low masses. Perhaps the most relevant negative feedback mechanism is heating due to photoionization (photoheating), since massive Pop~III stars produce vast amounts of ionizing photons (e.g. Tumlinson \& Shull 2000; Bromm, Kudritzki \& Loeb 2001b; Schaerer 2002), with escape fractions around unity (Whalen, Abel \& Norman 2004; Kitayama \& Yoshida 2005; Alvarez, Bromm \& Shapiro 2006a). Related to this, other radiative feedback mechanisms have also been considered (for a recent review, see Ciardi \& Ferrara 2005).

Of particular importance is the impact of radiation in the Lyman-Werner (LW) bands on neighbouring minihaloes, which might readily dissociate H$_2$ and suppress efficient cooling (e.g. Haiman, Rees \& Loeb 1997; Omukai \& Nishi 1999; Ciardi, Ferrara \& Abel 2000; Haiman, Abel \& Rees 2000; Machacek, Bryan \& Abel 2001; Ricotti, Gnedin \& Shull 2001, 2002a,b; Kitayama et al. 2001; Glover \& Brand 2001; Yoshida et al. 2003; Omukai \& Yoshii 2003; Mesinger, Bryan \& Haiman 2006; MacIntyre, Santoro \& Thomas 2006). Despite the extensive effort, no consensus has yet been reached on the overall significance of this effect. Alternatively, positive feedback due to an early X-ray background produced by relic black holes has been suggested (e.g. Haiman et al. 2000; Machacek, Bryan \& Abel 2003; Glover \& Brand 2003). Pop~III stars in the range $100$~--~$140$ and $260$~--~$500~\rmn{M}_{\odot}$ are expected to collapse directly to black holes after their main sequence lifetimes of $\la 3~\rmn{Myr}$ (Heger et al. 2003), emitting X-rays during the subsequent accretion of gas. Other possible sources include X-ray binaries and SN remnants (e.g. Natarajan \& Almaini 2000; Helfand \& Moran 2001). The buildup of an early X-ray background could enhance the production of free electrons in neighbouring minihaloes, thereby increasing the H$_2$ fraction and consequently the cooling rate. This would result in an increase of Pop~III star formation and could possibly compensate for negative feedback caused by photoheating or LW radiation.

Based on the revised value of the optical depth to Thomson scattering (e.g. Spergel et al. 2006), we find that Pop~III must have been terminated fairly rapidly (see Section~3). This leads us to concentrate on negative feedback, and in the following we will discuss the impact of the LW background, and subsequently of photoheating, on primordial star formation.

\subsubsection{Lyman-Werner Feedback}
The high masses and surface temperatures of Pop~III stars give rise to yields of the order of $10^{5}$ ionizing photons per stellar baryon (e.g Bromm et al. 2001b; Schaerer 2002). Naturally, the flux in the LW bands is also high, dissociating H$_{2}$ in neighbouring haloes and reducing the fraction of gas that can cool efficiently. To describe this feedback mechanism, we must find a connection between the background LW flux and the cooling rate of a halo, and also between the mass in stars and the associated photon yields.

For simplicity, we assume that all emitted LW photons escape into the IGM, and are thus available for H$_2$-dissociation in a nearby minihalo, but account for the redshifting of photons out of the LW bands. This approach clearly overestimates the LW background, yet will turn out to be instructive when discussing the results. The buildup of a homogeneous LW background proceeds over a very short period of time, since the comoving mean free path of a LW photon at $z=30$ is $\simeq 10$~Mpc (Mackey et al. 2003). Often, the presence of such a pervading background has been assumed in investigating the cooling process of a minihalo under LW irradiation. Specifically, Machacek et al. (2001) have derived an analytic fit to their simulation results, giving an estimate of the gas fraction that cools and reaches high densities as a function of halo mass and LW flux:
\begin{equation}
f_{\rmn{LW}}(M,J_{21})=0.06\,\rmn{ln}\left(\frac{M/\rmn{M}_{\odot}}{1.25\times 10^{5}+8.7\times 10^{5}F_{21}^{0.47}}\right)\mbox{\ ,}
\end{equation}
where $F_{21}=4\upi J_{21}$ is the flux in the LW bands, such that: $J_{\rmn{LW}}=J_{21}\times 10^{-21}~\rmn{erg}~\rmn{s}^{-1}~\rmn{cm}^{-2}~\rmn{Hz}^{-1}~\rmn{sr}^{-1}$. This formula explicitly shows that H$_{2}$ cooling rates rise towards higher halo masses, enabling them to withstand larger amounts of LW flux. Averaging this fraction over the halo mass range hosting Pop~III, and `normalizing' it to the value obtained for an absent LW background, yields the mass fraction of gas available for star formation as a function of $z$ and $J_{21}$:
\begin{equation}
\bar{f}_{\rmn{LW}}(J_{21},z)=\frac{\int_{M_{\rmn{min}}}^{M_{\rmn{max}}}\rmn{d}MMn_{\rmn{ST}}(M,z)f_{\rmn{LW}}(M,J_{21})}{\int_{M_{\rmn{min}}}^{M_{\rmn{max}}}\rmn{d}MMn_{\rmn{ST}}(M,z)f_{\rmn{LW}}(M,J_{21}=0)}\mbox{\ ,}
\end{equation}
where the normalisation accounts for the fact that Machacek et al. (2001) imply a definition for the star formation efficiency different from our choice of a constant $f_{*}$, independent of halo mass. Thus, the SFR at any given redshift is reduced relative to the no-feedback SFR of equation~(4) by precisely this factor:
\begin{equation}
\Psi_{*}'(J_{21},z)=\bar{f}_{\rmn{LW}}(J_{21},z)\Psi_{*}(z)\mbox{\ ,}
\end{equation}
where $J_{21}$ is implicitly a function of redshift as it is linked to the mass already incorporated into stars. To complete the feedback cycle, we must therefore connect the LW flux with the comoving density in stars, $\rho_{*}(z)$. For simplicity, we assume that the ratio of mean LW frequency to the relevant bandwidth is of the order of unity:
\begin{equation}
J_{\rmn{LW}}(z)\simeq\frac{hc}{4\upi m_{\rmn{H}}}\eta_{\rmn{LW}}\rho_{*}(z)\left(1+z\right)^{3}\mbox{\ ,}
\end{equation}
where $\eta_\rmn{LW}$ is the number of photons emitted in the LW bands per stellar baryon, and:
\begin{equation}
\rho_{*}(z)=\int_{z}^{z_{\rmn{r}}}\Psi_{*}(z')\left|\frac{\rmn{d}t}{\rmn{d}z'}\right|\rmn{d}z'\mbox{\ ,}
\end{equation}
such that $\Delta z=z_{\rmn{r}}-z$ corresponds to the maximum distance a LW photon can travel before it is redshifted out of the LW bands. Since all stellar populations contribute to the LW background, possibly with distinct values for $\eta_{\rmn{LW}}$, the star formation histories of Pop~II.5 and Pop~I/II are directly linked to Pop~III. Schaerer (2002) has provided time-integrated spectral properties of metal-free stars over a wide range of masses, and our calculations of $\eta_{\rmn{LW}}$ are based on this work. For Pop~III, we integrate over the Salpeter-weighted mass range $100$~--~$500~\rmn{M}_{\odot}$, which results in $\eta _{\rmn{LW}}\simeq 10^{4}$. As will be discussed in more detail later in this paper, Pop~II.5 is assumed to have masses $\ga 10~\rmn{M}_{\odot}$, leading us to adopt mass cut-offs at $10$ and $100~\rmn{M}_{\odot}$, and again a standard Salpeter slope. Using the spectra of Schaerer (2002), we find $\eta_{\rmn{LW}}\simeq 2\times 10^{4}$. For Pop~I/II, we adopt $\eta_\rmn{LW}=4\times 10^{3}$, which is comparable to the ionizing flux. The explicit SFRs used for Pop~II.5 and Pop~I/II correspond to the feedback parameter $K_{\rmn{w}}^{1/3}=2/3$, which will be discussed later. We note, however, that varying this parameter hardly affects our results.

Applying the above treatment allows us to describe Pop~III star formation in an evolving LW background. Our calculation is shown in Fig.~1 (dotted line), compared to the no-feedback case (solid line). Somewhat surprisingly, the overall amplitude of the effect is not too high, although we have conservatively assumed that no LW photons are lost. On the other hand, we have not included the effect of biasing since the build-up of a homogenuous LW background is very rapid. The shape of the modified SFR is also quite interesting, and can be explained by a decomposition into two competing processes: As the LW background increases, the Pop~III SFR tends to decline, yet as haloes become more massive, they can sustain higher LW fluxes, enabling ongoing Pop~III star formation towards lower redshifts. To summarize, LW feedback might not be as efficient as previously assumed (e.g. Haiman et al. 1997; Mackey et al. 2003), yet for definitive conclusions one must apply detailed numerical simulations that self-consistently solve the hydrodynamics and radiative transfer.

\begin{figure}
\begin{center}
\includegraphics[width=8cm,height=8cm]{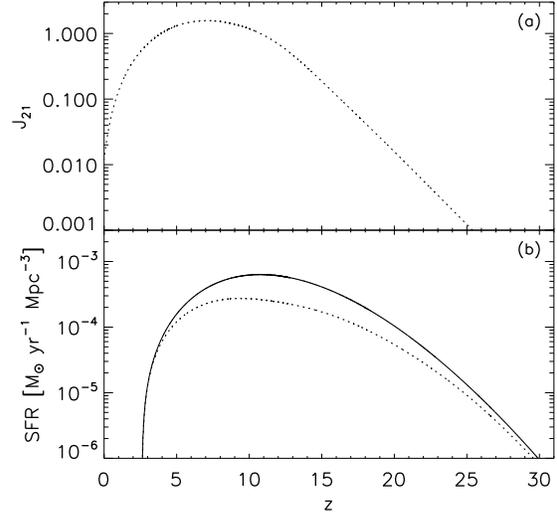}
\caption{(a) Buildup of the LW background in units of $10^{-21}~\rmn{erg}~\rmn{s}^{-1}~\rmn{cm}^{-2}~\rmn{Hz}^{-1}~\rmn{sr}^{-1}$. We here consider the LW radiation produced by all stellar populations. (b) The SFR for Pop~III, neglecting all feedback effects (solid line), and after inclusion of LW feedback (dotted line).}
\end{center}
\end{figure}

\subsubsection{Photoheating}
The emergence of the first H\,{\sc ii} regions around star-forming haloes has an important effect on high-redshift star formation. The temperature of ionized bubbles is raised to $\sim 10^{4}~\rmn{K}$, quenching star formation in affected regions, as primordial gas can no longer condense into the shallow gravitational wells of minihaloes. Since Pop~III stars produce copious amounts of ionizing photons, with escape fractions of the order of unity (e.g. Alvarez et al. 2006a), and haloes are strongly biased at these high redshifts, photoheating could in fact be a dominant feedback mechanism for Pop~III (e.g. Kramer \& Haiman 2006). To quantify this feedback, we begin with given ionization histories, and assume that Pop~III star formation is suppressed by a factor equal to the volume filling fraction of ionized regions, $Q_{\rmn{ion}}(z)$, weighted by the mean correlation between haloes. With this treatment, we might overestimate the strength of the feedback, as already-collapsed haloes are subject to photoevaporation rather than infall suppression due to photoheating, the former being a much slower process (e.g. Shapiro, Iliev \& Raga 2004; Iliev, Shapiro \& Raga 2005). However, since the collapse of neighboring minihaloes is synchronized only to within $\sim 10^{7}~\rmn{yr}$, compared to propagation times of the first H\,{\sc ii} regions of $\sim 10^{6}~\rmn{yr}$ (e.g. Alvarez et al. 2006a), we expect that most haloes have not yet reached high enough densities to be subject to photoevaporation.

To explicitly determine the effect of biasing, we use the well-known linear correlation function $\xi_{\rmn{hh}}(M,R,z)$ between two haloes of mass $M$ at a comoving distance $R$:
\begin{equation}
\xi_{\rmn{hh}}(M,R,z)=b^{2}(M,z)\xi_{\rmn{mm}}(R)\mbox{\ ,}
\end{equation}
where $b(M,z)$ is the linear bias of a halo of mass $M$ and $\xi_{\rmn{mm}}(R)$ is the mass correlation function (e.g. Mo \& White 2002). Since haloes are distributed according to the ST function, we can determine the average bias at any given redshift:
\begin{equation}
\bar{b}(z)=\frac{\int_{M_{\rmn{min}}}^{M_{\rmn{max}}}\rmn{d}MMb(M,z)n_{\rmn{ST}}(M,z)}{\int_{M_\rmn{min}}^{M_\rmn{max}}\rmn{d}MMn_{\rmn{ST}}(M,z)}\mbox{\ ,}
\end{equation}
which replaces $b(M,z)$ in equation~(10). The second step consists of linking the volume filling fraction of ionized regions to the typical size of an ionized bubble, which we do by assuming that the volume of a typical H\,{\sc ii} region is equal to $Q_{\rmn{ion}}(z)$ divided by the current number density of haloes, $n(z)$:
\begin{equation}
\frac{4}{3}\upi R^{3}(z)=\frac{Q_{\rmn{ion}}(z)}{{n}(z)}\mbox{\ .}
\end{equation}

With these preparations, we can ascertain the probability $p_{\rmn{ion}}(z)$ that a freshly collapsed halo lies within an already existing H\,{\sc ii} region:
\begin{equation}
p_{\rmn{ion}}(z)=Q_{\rmn{ion}}(z)\left[1+\xi_{\rmn{hh}}(z)\right]\mbox{\ ,}
\end{equation}
where the halo correlation function $\xi_{\rmn{hh}}(z)$ adopts the interpretation of an excess probability. This enables us to write the modified SFR with respect to the no-feedback SFR as:
\begin{equation}
\Psi_{*}'(z)=\left[1-p_{\rmn{ion}}(z)\right]\Psi_{*}(z)\mbox{\ .}
\end{equation}
To calculate the SFRs we must supply reionization histories, and choose Gaussian functions of the form $Q_\rmn{ion}(z)=\rmn{e}^{-\left(z-6\right)^{2}/w}$, where $w=57$, $25$ and $5$ corresponds to early, intermediate and late reionization, respectively. For all three cases, reionization ends at $z=6$, thus being in agreement with observations suggesting that the Universe was completely ionized by $z\ga 6$, e.g. the absence of the Gunn \& Peterson (1965) trough in the spectra of high-redshift quasars, and considerations of the thermal history of the IGM (e.g. Theuns et al. 2002; Hui \& Haiman 2003). Additionally, the three reionization histories reproduce optical depths of $\tau=0.12$, $0.09$ and $0.06$, thus spanning the current error bars of the {\it WMAP} measurement. For the three different values of $w$ as mentioned above, Panel~(a) in Fig.~2 shows early (dotted line), intermediate (dashed line) and late (dot-dashed line) reionization, whereas Panel~(b) traces the modified SFRs derived from these reionization histories. As is evident from the plots, the different epochs of reionization lead to vastly different SFRs, with efficiencies varying by more than one order of magnitude. This is primarily an effect of halo biasing, amplifying the impact of small volume filling fractions $Q_\rmn{ion}(z)$. We tentatively conclude that Pop~III ends very rapidly once a small number of minihaloes have formed stars.

At this point, it is important to note that we have not yet derived self-consistent reionization and star formation histories. However, it is instructive to outline the range of possibilities and defer conclusions about the most likely and self-consistent scenario to later sections.

\begin{figure}
\begin{center}
\includegraphics[width=8cm,height=8cm]{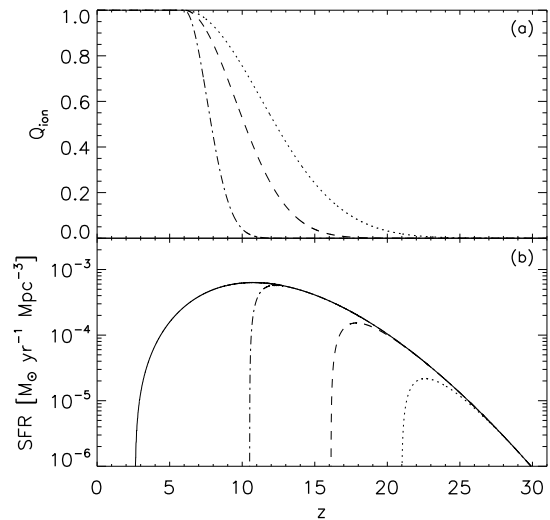}
\caption{(a) Early (dotted line), intermediate (dashed line) and late (dot-dashed line) reionization, here applied to calculate the suppression of Pop~III star formation by photoheating. (b) Comparison of the modified SFRs to the no-feedback case (solid line).}
\end{center}
\end{figure}

\subsection{Population~II.5}
The term Pop~III has traditionally been used to denote stars forming in extremely low metallicity environments, independent of the associated halo mass. In the following, we will deviate from this terminology, and instead distinguish between two populations of metal-free stars according to the environments in which they form. We thereby propose an intermediate stellar generation between Pop~III and Pop~I/II (see Mackey et al. 2003; Johnson \& Bromm 2006a).

As discussed in Section~2.1, haloes in the range $T_{\rmn{vir}}\simeq 10^{3}$~--~$10^{4}~\rmn{K}$ cool almost exclusively by molecular hydrogen transitions, and produce stars with an extremely top-heavy IMF. Yet gas assembled in more massive haloes goes through an ionized phase and experiences a different cooling process (Johnson \& Bromm 2006a,b). In this process, HD cooling allows the gas temperature to reach that of the cosmic microwave background (CMB) within a Hubble time. By reaching the CMB limit, these haloes produce stars with typical masses $\ga 10~\rmn{M}_{\odot}$, about an order of magnitude less than is the case for Pop~III (Johnson \& Bromm 2006a). Associating this intermediate generation of stars with metal-free haloes above $T_{\rmn{vir}}=10^{4}~\rmn{K}$ allows us to derive a SFR for Pop~II.5.

\subsubsection{Population~II.5 Star Formation}
To obtain a preliminary SFR for Pop~II.5, we once again apply the ST formalism, yet adjust the limits of integration when determining the collapsed mass fraction by setting the lower limit to $M_{\rmn{min}}=M_{\rmn{crit}}\left(T_{\rmn{min}}=10^{4}~\rmn{K}\right)$ and the upper limit to infinity. This satisfies the conditions for their formation sites, as these haloes go through an ionized phase in the course of their virialization (e.g. Johnson \& Bromm 2006a). The SFR is computed according to equation~(4), except that we use a different star formation efficiency, $f_{*}\sim 0.01$. This choice reflects the intermediate nature of Pop~II.5 between Pop~III and Pop~I/II, and will be further justified in later sections. The solid line in Panel~(b) of Fig.~3 shows the resulting SFR, implying that Pop~II.5 produces stars about an order of magnitude more efficiently than Pop~III (see Fig.~2), but also peaks at later times.

\subsubsection{Feedback Mechanisms}
Due to the deep potential wells of typical Pop~II.5 haloes ($\ga 10^{8}~\rmn{M}_{\odot}$), and the fact that cooling by atomic hydrogen sets in, photoheating is no longer a dominant feedback mechanism. Specifically, Dijkstra et al. (2004) have shown that suppression of gas infall is not as efficient in the high-redshift Universe as it is for $z\la 6$. We performed calculations including infall suppression in ionized regions of the IGM and verified this conclusion. Efficient self-shielding of ionized gas assembled in more massive haloes also renders LW feedback unimportant (see Johnson \& Bromm 2006b). Consequently, Pop~II.5 can only be terminated by chemical feedback, as multiple SNe in a host galaxy combine to form so-called `superwinds', expanding into the IGM and enriching neighbouring haloes with metals (e.g. Madau et al. 2001; Mori, Ferrara \& Madau 2002; Furlanetto \& Loeb 2003, 2005; Wada \& Venkatesan 2003). In the next section, we will therefore introduce a semianalytic model describing chemical enrichment, which subsequently leads to the termination of Pop~II.5.

\subsubsection{Chemical Feedback}
Once the metallicity of the IGM exceeds a critical level, the formation of primordial, metal-free stars is suppressed and star formation switches to a more conventional mode (e.g. Omukai 2000; Bromm et al. 2001a; Schneider et al. 2002). The precise value of this `critical metallicity' is still heavily debated, as some authors argue that dust production in the early Universe might be responsible for lowering $Z_{\rmn{crit}}$ to almost $10^{-6}~Z_{\odot}$ (e.g. Schneider et al. 2006, and references therein). However, due to remaining uncertainties we choose the more robust value $Z_{\rmn{crit}}=10^{-3.5}~Z_{\odot}$, which is based on the physics of fine structure cooling, and does not depend on whether dust is present or not, and if so, what its composition might be (e.g. Bromm et al. 2001a; Bromm \& Loeb 2003; Santoro \& Shull 2006). Pop~III stars in the range $140$~--~$260~\rmn{M}_{\odot}$ explode as Pair-Instability Supernovae (PISNe), with substantial metal yields $y\simeq 0.5$ (Heger et al. 2003), yet due to their rare nature they are likely inefficient polluters (also see Section~3.3). We therefore suggest that Pop~II.5 is largely responsible for enriching the IGM above the critical threshold and quenching metal-free star formation.

In modelling the propagation and distribution of metals, we closely follow the semianalytic treatment of SNe winds performed by Furlanetto \& Loeb (2005). Utilizing the well-known Sedov (1959) solution of an expanding bubble in a uniform background, they find an expression for the ratio of enriched mass to (total) halo mass:
\begin{eqnarray}
\eta (M)&=&27K_{\rmn{w}}f_{*}^{3/5}E_{51}^{3/5}\left(\frac{\omega_{\rmn{sn}}}{100~\rmn{M}_{\odot}}\right)^{-3/5}\nonumber\\
&&\times\left(\frac{M}{10^{10}~\rmn{M}_{\odot}}\right)^{-2/5}\left(\frac{1+z}{10}\right)^{-3/5}\mbox{\ ,}
\end{eqnarray}
where $K_{\rmn{w}}$ is a free parameter describing radiative losses, $E_{51}$ is the typical energy emitted by a SN in units of $10^{51}~\rmn{erg}$ and $\omega_{\rmn{sn}}$ is the mass in stars per SN ($\simeq 20~\rmn{M}_{\odot}$ for our assumed Pop~II.5 IMF). By integrating $\eta (M)$ over the ST function to determine the fraction of space with metals, and weighting it with the mean bias at a given reshift, Furlanetto \& Loeb derive a probability function $p_{\rmn{pris}}(z)$ for freshly collapsed gas to be pristine. We identify this quantity with the fraction of gas that has just collapsed and is available for Pop~II.5 star formation. In doing so, we implicitly assume that the mean metallicity of the fraction of space reached by an expanding superwind lies well above $Z_{\rmn{crit}}$. We will justify this assumption below.

The properties of these SNe follow directly from our assumed IMF. Stars in the range $\simeq 10$~--~$25~\rmn{M}_{\odot}$ explode as conventional core-collapse SNe with energies $E_{51}\simeq 1$, and metal yields $y\simeq 0.05$, while stars with masses above $\simeq 40~\rmn{M}_{\odot}$ collapse directly to black holes (Heger et al. 2003). The fate of stars with masses $25$~--~$40~\rmn{M}_{\odot}$ seems to be not so unambiguous, as those with low core angular momenta end their lives as faint SNe, while those with higher core angular momenta may explode as hypernovae (HNe), with energy outputs $E_{51}\ga 10$ (Heger et al. 2003). As a reasonable compromise, we therefore assume that all stars in the range $25$~--~$40~\rmn{M}_{\odot}$ explode as HNe with typical energies $E_{51}=10$, although variations herein do not significantly affect our results. Averaging the energy output and metal yield over the IMF results in $\bar{E_{51}}\simeq 2$ and $\bar{y}\simeq 0.04$.

With this information, we can determine the metal content of a region affected by a typical superwind. Explicitly, the metallicity of an enriched region is given by the ratio of mass in metals $M_{\rmn{z}}$ to the mass of the expanding bubble:
\begin{equation}
Z=\frac{\Omega_{\rmn{m}}}{\Omega_{\rmn{b}}}\frac{M_{\rmn{z}}}{\eta (M)M}\mbox{\ .}
\end{equation}
The mass in metals can be calculated from the total mass of the star-forming region by means of the star formation efficiency and the metal yield, resulting in:
\begin{eqnarray}
Z&=&\frac{y}{27K_{\rmn{w}}}f_{*}^{2/5}E_{51}^{-3/5}\left(\frac{\omega_{\rmn{sn}}}{100~\rmn{M}_{\odot}}\right)^{3/5}\nonumber\\
&&\times\left(\frac{M}{10^{10}~\rmn{M}_{\odot}}\right)^{2/5}\left(\frac{1+z}{10}\right)^{3/5}\mbox{\ .}
\end{eqnarray}
For a typical Pop~II.5 halo of $10^{8}~\rmn{M}_{\odot}$, forming at $z=15$, and with $K_{\rmn{w}}^{1/3}$ set to the fiducial value $2/3$, we find $Z\simeq 3\times 10^{-3}~Z_{\odot}$. This lies well above the critical metallicity and proves that our initial assumption was justified.

To obtain a SFR that accounts for enrichment by superwinds, we simply multiply the original SFR with $p_{\rmn{pris}}(z)$, thereby selecting the pristine fraction of freshly collapsed gas:
\begin{equation}
\Psi_{*}'(z)=p_{\rmn{pris}}(z)\Psi_{*}(z)\mbox{\ .}
\end{equation}
Using this semianalytic model, we compute star formation histories for three different radiation loss parameters, i.e. $K_{\rmn{w}}^{1/3}=1$, $2/3$ and $1/2$, corresponding to strong, intermediate and weak chemical feedback, respectively. Panel~(a) in Fig.~3 shows $p_{\rmn{pris}}(z)$ for $K_{\rmn{w}}^{1/3}=1$ (dotted line), $K_{\rmn{w}}^{1/3}=2/3$ (dashed line) and $K_{\rmn{w}}^{1/3}=1/2$ (dot-dashed line), whereas Panel~(b) shows the implied SFRs. As expected, higher radiative losses result in less chemical enrichment and more efficient star formation. Also, the termination process is much slower than is the case for Pop~III, due to the nature of the feedback: Wind hosts are heavily biased at these high redshifts, implying that enriching superwinds overlap and do not enrich large volume filling fractions, whereas the effect of biasing only marginally inhibits the propagation of ionization fronts.

\begin{figure}
\begin{center}
\includegraphics[width=8cm,height=8cm]{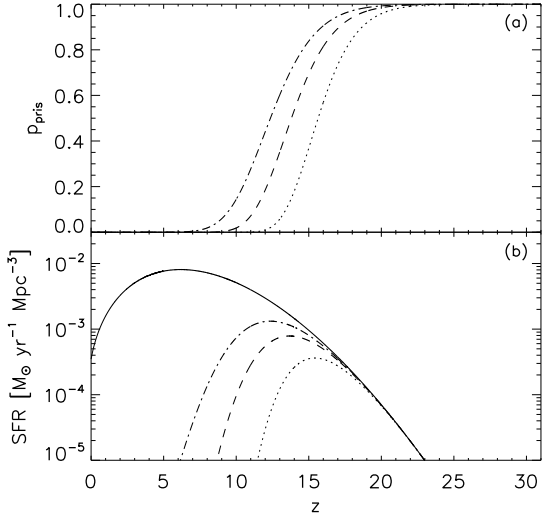}
\caption{(a) Evolution of $p_{\rmn{pris}}(z)$ for strong (dotted line), intermediate (dashed line) and weak (dot-dashed line) chemical feedback. (b) Comparison of the no-feedback SFR for Pop~II.5 (solid line) to the SFRs modulated by $p_{\rmn{pris}}(z)$.}
\end{center}
\end{figure}

\subsection{Population~I/II}
In order to compare the SFRs we have derived for Pop~III and Pop~II.5 to recent stellar generations, and also determine the reionization history of the Universe, we will briefly discuss our model for Pop~I/II. Closely following the treatment of Pop~II.5, and maintaining consistency in terms of our ST approach, we assume that haloes above $T_{\rmn{vir}}=10^{4}~\rmn{K}$ form Pop~I/II stars once they have been polluted by enriching superwinds:
\begin{equation}
\Psi_{*}'(z)=\left[1-p_{\rmn{pris}}(z)\right]\Psi_{*}(z)\mbox{\ ,}
\end{equation}
where $\Psi_{*}(z)$ is the no-feedback SFR introduced in Section~2.2.1. Fig.~4 shows the resulting SFR for Pop~I/II using $K_{\rmn{w}}=2/3$, compared to an accurate fit of recent star formation history by Springel \& Hernquist (2003) of the form:
\begin{equation}
\Psi_{*}(z)=K\frac{a\;\rmn{exp}\left[b\left(z-z_{\rmn{m}}\right)\right]}{a-b+b\;\rmn{exp}\left[a\left(z-z_{\rmn{m}}\right)\right]}\mbox{\ ,}
\end{equation}
where $K=0.15~\rmn{M}_{\odot}~\rmn{yr}^{-1}~\rmn{Mpc}^{-3}$, $a=14/15$, $b=3/5$ and $z_{\rmn{m}}=5.4$. Note that this expression was derived for the {\it WMAP} first year parameters, and the true SFR would be somewhat lower. In the redshift regime of interest, i.e. $z\la 15$, both plots are in good agreement, since at higher redshifts Pop~II.5 becomes the dominant mode.

In the next section, we will evaluate the SFRs we have derived for Pop~III, Pop~II.5 and Pop~I/II by comparing the ionizing photon and metal production to observational constraints.

\begin{figure}
\begin{center}
\includegraphics[width=8cm,height=5cm]{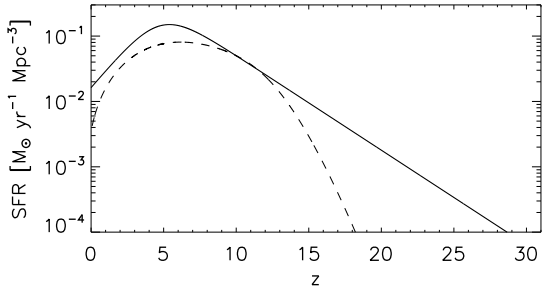}
\caption{Pop~I/II star formation for intermediate chemical feedback (dashed line), compared to an analytic fit by Springel \& Hernquist (solid line).}
\end{center}
\end{figure}

\section {Observational Signatures}
To test the validity of our models, we must compare our SFRs to observational signatures. This turns out to be a formidable task, as the sites of early star formation are the most distant sources of stellar radiation in the Universe, and all empirical constraints to date have been indirect. Yet the situation is rapidly improving by utilizing existing telescopes such as {\it Spitzer}, and pushing them to their limits (e.g. Barkana \& Loeb 2001; Bromm \& Larson 2004; Ciardi \& Ferrara 2005; Beers \& Christlieb 2005).

One of the most important clues on early star formation is the {\it WMAP} measurement of the optical depth to Thomson scattering, recently revised to be $\tau\simeq 0.09\pm 0.03$ (Spergel et al. 2006). By providing an integral constraint on the total ionizing photon production of the Universe, this measurement is well suited to validate our models. Related to this, we briefly consider the contribution of metal-free stars to the cosmic near infrared background (CNIRB), as an excess in the range $1$~--~$2~\umu\rmn{m}$ has been discovered after subtraction of all known foreground sources (Cambr\'{e}sy et al. 2001; Santos, Bromm \& Kamionkowski 2002).

Another observational probe with great promise concerns the distribution of metals in the Universe, as considerable improvements have been made in measuring the metallicity of the IGM by quasar absorption lines (e.g. Pettini 2001). By computing the mass fraction of metals ejected into the IGM, we compare our results to metal abundances of the Ly$\alpha$ forest (e.g. Songaila 2001), thereby elucidating the enrichment properties of Pop~III and Pop~II.5. Finally, we consider how metal-poor Galactic halo stars can be accomodated within our theoretical framework (for a recent review see Beers \& Christlieb 2005).

\subsection{Ionizing Photon Production}
The most significant change related to the release of the {\it WMAP} 3-yr results concerns the optical depth to Thomson scattering, as the refined measurement revealed a reduction by nearly $50$~per~cent. At first sight, this might imply that early star formation was greatly suppressed, yet due to revisions in other cosmological parameters as well (i.e. the fluctuation power at small scales), high-$z$ structure formation is rather shifted to lower redshifts by $\Delta z\simeq 5$, implying that primordial star formation is still relevant (e.g. Alvarez et al. 2006b). To investigate this connection more thoroughly, we must determine reionization histories and optical depths produced by our SFRs.

A straightforward manner of quantifying the evolution of H\,{\sc ii} regions is to assume that each photon escaping a host galaxy ionizes one intergalactic hydrogen or helium atom. Supplementing this by a term encompassing recombination of electrons results in a simple formula for the evolution of the volume filling fraction of ionized regions (e.g. Wyithe \& Loeb 2003, and references therein):
\begin{eqnarray}
\frac{\rmn{d}Q_{\rmn{ion}}(z)}{\rmn{d}z}&=&\frac{1}{n_{b}}\frac{\rmn{d}n_{\rmn{ion}}(z)}{\rmn{d}z}-\alpha_{\rmn{B}}n_{\rmn{b}}C(z)\nonumber\\
&&\times Q_{\rmn{ion}}^{2}(z)\left(1+z\right)^{3}\left|\frac{\rmn{d}t}{\rmn{d}z}\right|\mbox{\ ,}
\end{eqnarray}
where $n_{b}$ is the comoving density in baryons, $n_{\rmn{ion}}(z)$ the comoving density in ionizing photons, $\alpha_{\rmn{B}}$ the case B recombination coefficient and $C(z)$ the clumping factor. This method is well established in modelling the evolution of H\,{\sc ii} regions, and we refer to Wyithe \& Loeb (2003) for details. Throughout this paper, we choose $\alpha_{\rmn{B}}=2.6\times10^{-13}~\rmn{cm}^{3}~\rmn{s}^{-1}$, assuming that ionized regions are photoheated to $\sim 10^{4}~\rmn{K}$ (e.g. Osterbrock \& Ferland 1989). To concentrate on the key effects, we make the following simplifying assumptions: (i) We neglect ionizing photons emitted by quasars. (ii) We do not distinguish between helium and hydrogen, implying that helium can only be ionized once. (iii) We use a simple analytic fit of the form $C(z)=1+9\left[\left(1+z\right)/7\right]^{-2}$, with $C(z)=\rmn{const.}=10$ for $z<6$, to describe the evolution of the clumping factor. This is a compromise between contemporary approaches, and we refer to Haiman \& Bryan (2006) for further discussion.

The crucial quantity in equation~(21) is $n_{\rmn{ion}}(z)$, as it links our SFRs to the ionizing photon production. We establish this connection by assuming that the number of ionizing photons emitted per redshift interval is proportional to the SFR. Explicitly, the conversion from newly formed mass in stars to the comoving density of ionizing photons is given by:
\begin{equation}
\frac{1}{n_{b}}\frac{\rmn{d}n_{\rmn{ion}}(z)}{\rmn{d}z}=\frac{1}{\rho_{\rmn{m}}}\frac{\Omega_{\rmn{m}}}{\Omega_{\rmn{b}}}f_{\rmn{esc}}
\eta_{\rmn{ion}}\Psi_{*}(z)\left|\frac{\rmn{d}t}{\rmn{d}z}\right|\mbox{\ ,}
\end{equation}
where $f_{\rmn{esc}}$ is the escape fraction of ionizing photons from the host halo and $\eta_{\rmn{ion}}$ the number of ionizing photons emitted per stellar baryon. Evidently, $f_{\rmn{esc}}$ and $\eta_{\rmn{ion}}$ have different values for each population, and much effort has been undertaken to constrain these parameters. Specifically, Alvarez et al. (2006a) have found the escape fraction of ionizing photons in minihaloes to be between $0.5$ and $1$, leading us to adopt $f_{\rmn{esc}}=0.7$ for Pop~III. The escape fraction for typical Pop~I/II galaxies is a strong function of redshift (e.g. Inoue, Iwata \& Deharveng 2006), yet for early galaxies a reasonable estimate is $0.1$, whereas for Pop~II.5 we can only guess that the best-fitting value must lie between $0.7$ and $0.1$. We hereby use $f_{\rmn{esc}}\simeq 0.3$, but keep in mind that it is generally a free parameter. In deriving $\eta_{\rmn{ion}}$ for Pop~III and Pop~II.5, we proceed similarly to Section~2, and use the spectra for metal-free stars provided by Schaerer (2002). This yields $\eta_{\rmn{ion}}\simeq 9\times 10^{4}$ for Pop~III, and $\eta_{\rmn{ion}}\simeq 3\times 10^{4}$ for Pop~II.5. A common estimate for the ionizing photon production of Pop~I/II is $\eta_{\rmn{ion}}\simeq 4\times 10^{3}$.

With these quantities at hand, we compute the ionization histories $Q_{\rmn{ion}}(z)$ implied by our SFRs. Panel~(a) in Fig.~5 shows the ionization histories for the three distinct SFRs for Pop~III determined in Section~2, i.e. early (dotted line), intermediate (dashed line) and late (dot-dashed line) reionization, combined with intermediate chemical feeback for Pop~II.5 (although variations in $K_{\rmn{w}}$ hardly affect our results). By comparing Fig.~5 with Fig.~2, we conclude that intermediate reionization is by far the most likely and self-consistent model, as especially late reionization permits a genuinely high SFR (see Fig.~2), which induces a large contribution to the ionizing photon production, and ultimately contradicts the proposed reionization history. This further confirms our assumption that Pop~III must be terminated fairly rapidly.

In a final step, we determime the optical depth to Thomson scattering implied by these ionization histories:
\begin{equation}
\tau =c\sigma_{\rmn{T}}n_{b} \int_{0}^{z}dz'\,Q_{\rmn{ion}}(z')\left(1+z'\right)^{3}\left|\frac{\rmn{d}t}{\rmn{d}z'}\right|\mbox{\ ,}
\end{equation}
where $z$ is the redshift of emission and $\sigma_{\rmn{T}}$ the cross-section to Thomson scattering (e.g. Wyithe \& Loeb 2003). According to this equation, the optical depth of a given ionization history is a well determined quantity, and Panel~(b) in Fig.~5 shows the optical depths corresponding to the three reionization histories of Panel~(a). Not surprisingly, late reionization (as in Fig.~2) produces a value near the upper limit, whereas for early and intermediate reionization the resulting optical depth is very close to $\tau =0.09$.

At this point, we have also tested the validity of our star formation efficiencies for Pop~III and Pop~II.5. For values much higher than our fiducial choices (see Table~1), we find optical depths that are unrealistically large. On the other hand, much smaller values completely neglect primordial star formation for ionization purposes. Taken together, this leads us to maintain the values we assumed in Section~2.

\begin{figure}
\begin{center}
\includegraphics[width=8cm,height=8cm]{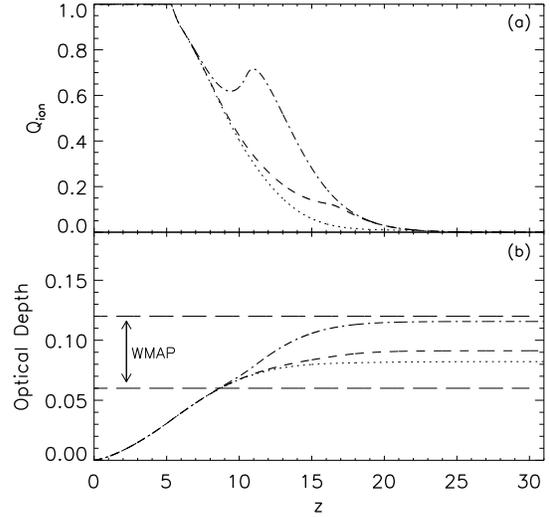}
\caption{Ionization histories (a), and implied optical depths (b) for the cases of preliminary early (dotted line), intermediate (dashed line) and late (dot-dashed line) reionization, as used to calculate the SFRs for Pop~III (also see Fig.~2). For comparison, Panel~(b) includes the range of allowed values for the optical depth as measured by {\it WMAP} (long-dashed lines).}
\end{center}
\end{figure}

\subsection{Near Infrared Background}
The high ionizing photon yields of typical Pop~III and Pop~II.5 stars might imply a significant contribution to the CNIRB. This possibility was investigated in detail by Santos et al. (2002), as an excess of the order $\sim 1~\rmn{nW}~\rmn{m}^{-2}~\rmn{sr}^{-1}$ in the wavelength regime $1$~--~$2~\umu\rmn{m}$ was detected by the Diffuse Infrared Background Experiment (DIRBE) on the {\it Cosmic Background Explorer} ({\it COBE}; e.g. Cambr{\'e}sy et al. 2001). The possible contribution to this wavelength range is made up of Ly$\alpha$ photons, redshifted by the amount $\Delta z\simeq 7$ on the lower end, and $\Delta z\simeq 15$ on the higher end, implying that this period of time was relevant for producing the observed excess. Since Pop~III is already extinct at these redshifts (we only consider intermediate reionization), Pop~II.5 is the only remaining candidate. To determine its importance for the CNIRB, we link the ionizing photon production to the mass in stars, and for simplicity assume that all ionizing photons are reprocessed through recombinations into Ly$\alpha$ emission:
\begin{equation}
J_{\rmn{NIR}}\simeq\frac{hc}{4\upi m_{\rmn{H}}}\eta_{\rmn{ion}}\int_{7}^{15}\Psi_{*}(z)\left|\frac{\rmn{d}t}{\rmn{d}z}\right|\rmn{d}z\mbox{\ .}
\end{equation}
Here, we have once again assumed that $\bar{\nu}_{\rmn{NIR}}/\Delta\nu_{\rmn{NIR}}$ is of the order of unity. Fairly independent of $K_{\rmn{w}}$, this results in an upper limit of $J_{\rmn{NIR}}\sim 10^{-3}~\rmn{nW}~\rmn{m}^{-2}~\rmn{sr}^{-1}$, which is much too low to account for the observed discrepancy. However, this result does not exclude the possibility of tracing primordial stars with fluctuations of the cosmic infrared background, as suggested by Kashlinsky et al. (2004).

\subsection{Metal Production}
In recent years, detailed studies of quasar absorption lines have provided a means of probing metal abundances of Ly$\alpha$ systems with column densities as low as $10^{14}~\rmn{cm}^{-2}$, made possible by improvements in data analysis techniques (e.g. Pettini 2001, and references therein). One of the most important developments concerns the discovery of a metallicity floor in the Ly$\alpha$ forest of the order $10^{-3}~Z_{\odot}$ in the redshift range $z\simeq 2$~--~$6$, which cannot be explained by galactic outflows of normal Pop~I/II galaxies (e.g. Songaila 2001; Aguirre et al. 2005). This implies that early star formation might have been responsible for establishing such a homogeneous level of enrichment. Following up on this conclusion, we discuss enrichment models for both Pop~III and  Pop~II.5, and elucidate their contributions to the observed metal abundances of the IGM.

\subsubsection{Intergalactic Enrichment by Pop~III}
As Heger et al. (2003) have discussed, Pop~III stars in the range $140$~--~$260~\rmn{M}_{\odot}$ end their lives as PISNe, extremely violent SNe explosions with energies $E_{51}\la 100$ and metal yields $y\la 0.5$. Due to the shallow potential wells of minihaloes, these massive explosions completely evacuate their gaseous content, and distribute metals into the IGM with high efficiency and homogeneity (e.g. Bromm et al. 2003; Yoshida et al. 2004). On the other hand, Pop~III stars with initial masses in the range $100$~--~$140$ and $260$~--~$500~\rmn{M}_{\odot}$ collapse directly to black holes, making them negligible for enrichment purposes (Heger et al. 2003). To find the correctly weighted metal yield, we average $y$ over the IMF, resulting in $\bar{y}\simeq 0.2$. This quantity is essential in modelling the enrichment process by Pop~III, as it connects the mass in stars to the mass in metals. Applying the argument of homogeneous enrichment, we estimate a lower limit of the IGM metallicity by determining the ratio of density in metals to the background density for a given SFR:
\begin{equation}
Z_{\rmn{IGM}}(z)=\frac{\bar{y}}{\rho_{\rmn{m}}}\int_{z}^{\infty}\Psi_{*}(z')\left|\frac{\rmn{d}t}{\rmn{d}z'}\right|\rmn{d}z'\mbox{\ .}
\end{equation}

Fig.~6 shows the resulting metallicity of the IGM for intermediate reionization (dashed line), compared to the fiducial value $Z\sim 10^{-3}~Z_{\odot}$ for measurements of the Ly$\alpha$ forest (e.g. Songaila 2001), and the critical metallicity $Z_{\rmn{crit}}=10^{-3.5}~Z_{\odot}$ (long-dashed lines). For the case of highly efficient mixing, chemical enrichment by Pop~III does not raise the globally averaged metallicity above $Z_{\rmn{crit}}$, implying that Pop~II.5 star formation remains possible even in regions affected by PISNe winds. As is evident from Fig.~6, we predict a minimum bedrock metallicity in the IGM of the order $Z_{\rmn{IGM}}\sim 10^{-5}~Z_{\odot}$, much lower than current measurements of the Ly$\alpha$ forest imply. This is not necessarily a contradiction, as the lowest column density regions of the IGM are not yet accessible for probing. Detecting such a low bedrock metallicity would provide a telltale signature of Pop~III star formation in early minihaloes.

\begin{figure}
\begin{center}
\includegraphics[width=8cm,height=5cm]{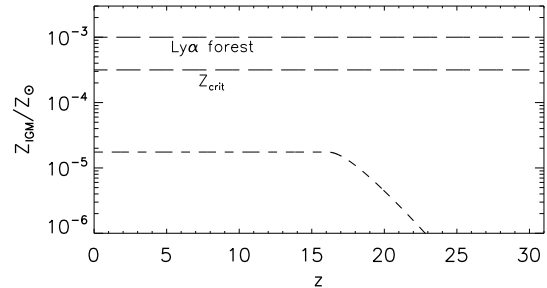}
\caption{Chemical Enrichment of the IGM by Pop~III for intermediate reionization (dashed line), compared to the Ly$\alpha$ forest and the critical metallicity $Z_{\rmn{crit}}$ (long-dashed lines).}
\end{center}
\end{figure}

\subsubsection{Local Enrichment by Pop~II.5}
The assumption of homogeneous metal distribution throughout the IGM breaks down for Pop~II.5, as typical explosion energies of conventional core-collapse SNe are significantly lower than those of PISNe, and the Universe at large has grown much more inhomogeneous since the epoch of Pop~III. We therefore assume that metals are distributed only inside enriching superwinds, and use equation~(15) to determine the average density of such an enriched region:
\begin{equation}
\bar{\rho}(z)=\int_{M_{\rmn{min}}}^{\infty}\rmn{d}MM\eta (M,z)n_{\rmn{ST}}(M,z)\mbox{\ .}
\end{equation}
Inserting $\bar{\rho}$ instead of $\rho_{\rmn{m}}$ into equation~(25), and adopting the relevant values of $\bar{y}$, $E_{51}$ and $f_{*}$ for Pop~II.5, we determine the average metallicity of regions affected by superwinds. For our fiducial parameter $K_{\rmn{w}}=2/3$ this yields $Z\simeq 2\times 10^{-3}~Z_{\odot}$, which is in very good agreement with our calculation of Section~2, and lies close to measurements of low column density Ly$\alpha$ clouds (e.g. Songaila 2001). The metallicity exceeds $Z_{\rmn{crit}}$ by almost one order of magnitude, further confirming our initial assumption that superwinds quench metal-free star formation in affected regions. An intuitively unexpected result is that stronger chemical feedback results in less efficient metal production, which can be explained by the fact that a decrease in radiative losses enables the enriching winds to pollute larger regions of the IGM, thereby terminating Pop~II.5 more efficiently.

\subsubsection{Metal-Poor Halo Stars}
The discovery of metal-poor stars in the halo of the Milky Way with extremely low iron content, yet excess light element abundances, has triggered discussion on the properties of the first stars (e.g. Beers \& Christlieb 2005). Various authors have argued that a less top-heavy IMF is more successfull in reproducing the observed abundance patterns, as faint SNe and HNe of intermediate mass stars produce metal yields consistent with those found in metal-poor halo stars, provided these are of second-generation and formed from the relics of a metal-free population (e.g. Heger \& Woosley 2002; Umeda \& Nomoto 2002, 2003, 2005; Tumlinson et al. 2004; Daigne et al. 2004; Iwamoto et al. 2005). Based on the detailed physics of gas cooling, we have found primordial star formation to be dominated by stars $\ga 10~\rmn{M}_{\odot}$, and can therefore naturally account for the existence of metal-poor halo stars.

\section{Summary and Conclusions}
We have constructed star formation histories for two distinct populations of metal-free stars, characterized by their respective IMFs and formation environments. Specifically, we have attributed Pop~III to form in minihaloes with characteristic stellar masses $\ga 100~\rmn{M}_{\odot}$, whereas second-generation haloes of $\ga 10^{8}~\rmn{M}_{\odot}$ give rise to an intermediate Population~II.5, enabling the formation of $\ga 10~\rmn{M}_{\odot}$ stars by activating HD as an important low-temperature coolant. Acknowledging that both populations are subject to radiative and chemical feedback, we have shown that Pop~III is terminated in the course of reionization, whereas Pop~II.5 ends with the distribution of metals into the IGM by enriching superwinds, raising the metallicity in affected regions above $Z_{\rmn{crit}}=10^{-3.5}~Z_{\odot}$. We have confirmed the validity of our models by determining reionization histories and optical depths, and have excluded a dominant primordial contribution to the excess of the CNIRB background in the wavelength regime $1$~--~$2~\umu\rmn{m}$. Finally, we have discussed chemical enrichment of the IGM by Pop~III and Pop~II.5, finding that Pop~III produces a minimum bedrock metallicity in the IGM of the order $Z_{\rmn{IGM}}\sim 10^{-5}~Z_{\odot}$, whereas Pop~II.5 enriches the IGM to levels of the low column density Ly$\alpha$ forest. Summarizing these results, Fig.~7 compares the best-fitting SFRs for all three stellar populations, whereas Table~1 shows their characteristic quantities.

\begin{figure}
\begin{center}
\includegraphics[width=8cm,height=5cm]{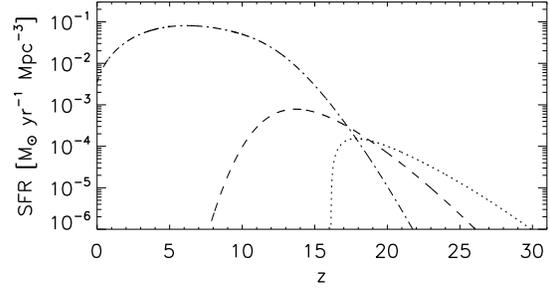}
\caption{Three epochs of cosmological star formation history: Pop~III (dotted line), Pop~II.5 (dashed line) and Pop~I/II (dot-dashed line).}
\end{center}
\end{figure}

\begin{table}
\begin{center}
\begin{tabular}{lccc} \hline
& Pop~III & Pop~II.5 & Pop~I/II \\ \hline
$M_{\rmn{char}}$ & $\ga 100~\rmn{M}_{\odot}$ & $\ga 10~\rmn{M}_{\odot}$ & $\sim 1~\rmn{M}_{\odot}$ \\
$T_{\rmn{min}}$ & $10^{3}$ K & $10^{4}$ K & $10^{4}$ K \\
$f_{*}$ & $0.001$  & $0.01$ & $0.1$ \\
$\eta_{\rmn{ion}}$ & $9\times 10^{4}$  & $3\times 10^{4}$ & $4\times 10^{3}$ \\
$f_{\rmn{esc}}$ & $0.7$  & $0.3$ & $0.1$ \\
$\bar{E}_{51}$ & $20$ & $2$ & $0.05$ \\
$\bar{y}$ & $0.2$ & $0.04$ & $0.005$ \\ \hline
\end{tabular}
\caption{Characteristic quantities of Pop~III, Pop~II.5 and Pop~I/II, from top to bottom: Typical stellar mass, minimum halo virial temperature, star formation efficiency, ionizing photon production, escape fraction, IMF-averaged explosion energy and IMF-averaged metal yield.}
\end{center}
\end{table}

The introduction of an intermediate generation of metal-free stars by means of formation environment provides a possible physical explanation to the empirically derived argument that early star formation might have been dominated by a less top-heavy IMF (e.g. Umeda \& Nomoto 2002, 2003; Tumlinson et al. 2004; Tumlinson 2006a,b). One hint comes from the realization that a single, top-heavy primordial population might overproduce metals while underproducing ionizing photons (e.g. Ricotti \& Ostriker 2004; Daigne et al. 2004, 2005; Matteucci \& Calura 2005). Some authors have also pointed out that metal yields of low column density Ly$\alpha$ clouds at $z\simeq 2$~--~$6$ show enrichment signatures by more intermediate mass stars (e.g. Daigne et al. 2004, 2005; Matteucci \& Calura 2005). Another argument focuses on the abundance patterns of metal-poor Galactic halo stars, as they indicate prior enrichment by conventional core-collapse SNe instead of PISNe (e.g. Heger \& Woosley 2002; Umeda \& Nomoto 2002, 2003, 2005; Tumlinson et al. 2004). By additionally considering the chemical evolution of the Galaxy, it has been found that a more moderate IMF dominated by $\ga 10~\rmn{M}_{\odot}$ stars best fits the currently available observational constraints (e.g. Tumlinson 2006a,b). Most recently, the revision of the optical depth to Thomson scattering has further weakened the case for a purely top-heavy primordial population, even in the light of a significantly reduced amplitude of the matter fluctuation power (e.g. Choudhury \& Ferrara 2006; Haiman \& Bryan 2006). Taken together, these arguments provide a strong case for a more moderate IMF, yet thus far no physical framework has been found to accomodate such lower mass stars. In this paper, we have shown that the physics of HD cooling lead to a natural transition in terms of typical stellar masses, and force a distinction between two populations of metal-free stars.

What was the relative importance of Pop~III and Pop~II.5? From our models, we estimate that only $\simeq 10$~per~cent of all mass involved in metal-free star formation went into Pop~III, and that Pop~II.5 was the dominant, and typical, mode of primordial star formation. This might suggest that Pop~III had a negligible impact on observable quantities, yet due to their extreme nature they might have nevertheless left distinct signatures. Especially their influence on the IGM could have been notable, as we expect that only PISNe are able to enrich the lowest column density regions in a homogeneous manner. Another clue might be provided by a characteristic reionization signature, as Pop~III stars have unique spectra (e.g. Tumlinson \& Shull 2000; Tumlinson, Giroux \& Shull 2001; Bromm et al. 2001b; Oh, Haiman \& Rees 2001; Tumlinson, Shull \& Venkatesan 2003; Venkatesan, Tumlinson \& Shull 2003). In light of these findings, it is essential to push observational frontiers to ever higher redshifts, yet also extend the reach of numerical simulations to more complex cosmological environments. In future work, we plan to apply such simulations to address the issues of intergalactic metal enrichment and star formation in atomic cooling haloes.

\section{Acknowledgments}
We would like to thank Marcelo A. Alvarez, Jarrett L. Johnson and Eiichiro Komatsu for fruitful discussions. We are grateful to Zoltan Haiman, Jason Tumlinson and the anonymous referee for their helpful comments.

\end{document}